# The Euclid mission design
# [9904-19]

Giuseppe D. Racca[*,a], René Laureijs[a], Luca Stagnaro[a], Jean-Christophe Salvignol[a], José Lorenzo Alvarez[a], Gonzalo Saavedra Criado[a], Luis Gaspar Venancio[a], Alex Short[a], Paolo Strada[a], Tobias Bönke[a], Cyril Colombo[a], Adriano Calvi[a], Elena Maiorano[a], Osvaldo Piersanti[a], Sylvain Prezelus[a], Pierluigi Rosato[a], Jacques Pinel[a], Hans Rozemeijer[a], Valentina Lesna[a], Paolo Musi[b], Marco Sias[b], Alberto Anselmi[b], Vincent Cazaubiel[c], Ludovic Vaillon[c], Yannick Mellier[d,g], Jérôme Amiaux[d], Michel Berthé[d], Marc Sauvage[d], Ruyman Azzollini[h], Mark Cropper[h], Sabrina Pottinger[h], Knud Jahnke[i], Anne Ealet[j], Thierry Maciaszek[k], Fabio Pasian[l], Andrea Zacchei[l], Roberto Scaramella[m], John Hoar[e], Ralf Kohley[e], Roland Vavrek[e], Andreas Rudolph[f], Micha Schmidt[f]

[a]European Space Agency / ESTEC, Keplerlaan 1, 2201 AZ Noordwijk, The Netherlands
[b]Thales Alenia Space Italia, Strada Antica di Collegno 253, 10146, Torino, Italy
[c] Airbus Defense & Space, 31 rue des cosmonautes, 31402 Toulouse, Cedex France
[d]IRFU, Service d'Astrophysique, CEA Saclay, F-91191 Gif-sur-Yvette Cedex, France
[e]European Space Agency / ESAC, Villanueva de la Cañada, E-28692 Madrid, Spain
[f]European Space Agency / ESOC, Robert-Bosch-Str. 5, 64293 Darmstadt, Germany
[g]CNRS-UPMC, Institut d'Astrophysique de Paris, 98B Bd Arago, F-75014 Paris
[h]Mullard Space Science Laboratory, University College London, Holmbury St Mary, Dorking, Surrey RH5 6NT, UK
[i]Max Planck Institute for Astronomy, Koenigstuhl 17, D-69117, Heidelberg, Germany
[j]Centre de Physique des Particules de Marseille, Aix-Marseille Université, CNRS/IN2P3, F-13288 Marseille, France
[k]Centre National d'Etudes Spatiales, 18 avenue Edouard Belin, F-31401 Toulouse Cedex 9, France
[l]INAF-Osservatorio di Trieste, Via G.B. Tiepolo 11, I-31131 Trieste, Italy
[m]INAF-Osservatorio di Roma, Via Frascati 33, I-00040 Monteporzio Catone (Roma), Italy

## ABSTRACT

Euclid is a space-based optical/near-infrared survey mission of the European Space Agency (ESA) to investigate the nature of dark energy, dark matter and gravity by observing the geometry of the Universe and on the formation of structures over cosmological timescales. Euclid will use two probes of the signature of dark matter and energy: Weak gravitational Lensing, which requires the measurement of the shape and photometric redshifts of distant galaxies, and Galaxy Clustering, based on the measurement of the 3-dimensional distribution of galaxies through their spectroscopic redshifts. The mission is scheduled for launch in 2020 and is designed for 6 years of nominal survey operations. The Euclid Spacecraft is composed of a Service Module and a Payload Module. The Service Module comprises all the conventional spacecraft subsystems, the instruments warm electronics units, the sun shield and the solar arrays. In particular the Service Module provides the extremely challenging pointing accuracy required by the scientific objectives. The Payload Module consists of a 1.2 m three-mirror Korsch type telescope and of two instruments, the visible imager and the near-infrared spectro-photometer, both covering a large common field-of-view enabling to survey more than 35% of the entire sky. All sensor data are downlinked using K-band transmission and processed by a dedicated ground segment for science data processing. The Euclid data and catalogues will be made available to the public at the ESA Science Data Centre.

---

[*] Euclid Project Manager, giuseppe.racca@esa.int; phone +31 71 565 4618; fax: F +31 71 565 5244; http://sci.esa.int/euclid

Starting from the overall mission requirements, we describe the spacecraft architectural design and expected performance and provide a view on the current project status.

## 1. INTRODUCTION AND SCIENCE CASE

In June 2012 the Science Programme Committee of ESA selected for implementation the Euclid mission as the next Medium Class mission of the ESA Cosmic Vision 2015-2025. Euclid is a space-based optical/near-infrared survey mission designed to investigate the nature of dark energy, dark matter and gravity by observing their signatures on the geometry of the Universe [1].

This paper is organised as follows. This section introduces the mission and summarises its science case. Section 2 describes the mission architecture, launch and survey design. Section 3 gives an overview on the spacecraft design comprising a service module (SVM) a payload module (PLM) and its instruments. Section 4 describes the Ground Segment, divided into Science and Operational ground segment. Section 5 gives a summary of the mission performance as estimated at the Mission Preliminary Design Review (MPDR) and section 6 will finally provide the programmatic status of the project.

Euclid is ESA's next cosmology mission after Planck [2]. While Planck mapped the structure and the properties of the early Universe at a redshift of $z\sim1100$, Euclid will map the evolution of cosmic structures from $z\sim2$ until now. Euclid uses the clustering of matter, which includes both dark and luminous matter, to measure the accelerated expansion of the Universe at different cosmological times. The accelerated expansion is attributed to a substance of unknown nature, dubbed Dark Energy (DE) [3]. Euclid aims at measuring the equation of state parameter $w(z)$, relating the DE 'fluid' pressure with its density, to an accuracy of 2% for the constant component, and 10% for a possible variation as a function of redshift. If $w(z)=$ constant $=-1$, then the universe can be described by General Relativity with an additional cosmological constant providing the accelerated expansion and Cold Dark Matter for the mass ($\Lambda$CDM). However at the moment such a description cannot be reconciled with the standard model of particle physics. A significant non-zero variation would mean that either a new "dark energy" component with exotic physical properties exists in the Universe or our understanding of General Relativity needs to be revisited. It is the accuracy of the Euclid measurements that will put strong constraints on any fundamental physics theory [4].

Euclid will use a number of cosmological probes to measure the clustering properties but is optimised for two methods: (1) Galaxy Clustering (GC): measurement of the redshift distribution of galaxies from their H$\alpha$ emission line survey using near-infrared slitless spectroscopy and (2) Weak Lensing (WL): measurement of the distortion of the galaxy shapes due to the gravitational lensing caused by the - predominantly dark - matter distribution between distant galaxies and the observer. The resulting galaxy shear field can be transformed into the matter distribution. This is done in a number of redshift bins to derive the expansion of the dark matter as a function of redshift. In addition, GC provides also direct information of the validity of General Relativity because we can monitor the evolution of structures subject to the combined effects of gravity, which forces clumping of matter, and the opposing force caused by the accelerated expansion. GC maps the distribution of the luminous, baryonic matter whereas WL measures the properties of the combination of both luminous and dark matter. The complementarity of the two probes will provide important additional information on possible systematics, which limit the accuracy of each of the probes.

The mission will address the following items and associated key questions:

- Is the Dark Energy simply a cosmological constant, or is it a field that evolves dynamically with the expansion of the Universe?

- Alternatively, is the apparent acceleration instead a manifestation of a breakdown of General Relativity on the largest scales, or a failure of the cosmological assumptions of homogeneity and isotropy?

- What is Dark Matter? What is the absolute neutrino mass scale and what is the number of relativistic species in the Universe as opposed to the structures predicted by the cold (non-relativistic) dark matter component?

- What are the initial conditions after the Big Bang, the power spectrum of primordial density fluctuations, which seeded large-scale structure, and are they described by a Gaussian probability distribution?

In order to accomplish the above science objectives, Euclid must survey a large fraction of the sky: 15,000 deg$^2$, or 36% of the celestial sphere and image billions of galaxies out to *z~2* up to an AB magnitude of 24.5 for 10σ extended objects in the visible band and of 24 for 5σ point sources in the NIR bands with the optical quality expressed by Table 6 and Table 7. The mission lifetime has been set to 6 years (6 months margin included), hence a telescope with a stable, large field of view is required.

## 2. MISSION ARCHITECTURE

The Euclid mission architecture is strongly driven by the science requirements and programmatic constraints. Major issues of the sky survey are its speed, depth, precision, and imaging quality while the main programmatic constraint is the mission duration. The survey speed is guaranteed by the combination of a large field of view, about 0.54 deg$^2$, and an optimised survey strategy. Ensuring the high image quality leads to demanding requirements on the pointing and thermo-elastic stability. The survey depth leads to a minimum telescope aperture, dedicated baffling design, low temperature optics and detectors, a cold telescope for low near-infrared background and on-board data processing for the noise reduction of the near-infrared detectors.

A large amplitude orbit around the second Sun-Earth Lagrange point (SEL2) has been selected because it imposes minimum constraints on the observations and allows scanning of the sky outside the galactic latitude *b* ±30 degree band around the Milky Way within the mission duration. The Euclid spacecraft will be launched from the Guiana Space Centre, Kourou, on board a Soyuz ST 2.1-B. The launch date and time determine the ellipticity and size of the operational orbit and influence the Sun-Spacecraft-Earth angle plus the daily visibility from the ground station. The launch is possible in most of the days of the year with minor restrictions to avoid eclipses during transfer and in the operational orbit, and by the angle between the Sun and telescope aperture during transfer. Once in operational orbit the spacecraft performs a step-and-stare scanning of the sky.

### 2.1 Survey design

Euclid aims to cover very large areas with great stability, thermal stresses must be minimised and this impacts on operations. In fact, the allowed pointing range is limited to observe orthogonally to the Sun, in a range of -3º towards to +10º from the orthogonal. In practice Euclid can scan parts of a circle on the sky along the ecliptic meridian; the visibility at ecliptic equator is ~ one week per semester (the target can be seen six months later along the same circle). This visibility period increases with the distance from the equator, up to two small circles at the ecliptic poles, which have perennial visibility.

The elementary observation sequence of a field is composed of four frames of the 0.54 deg$^2$ common area, observed with a dither step in-between. During each frame the visual instrument (VIS) and the Near Infrared Spectro-Photometer (NISP) spectrometer carry out exposures of the sky simultaneously. Subsequently, because of the disturbing vibration from filter wheel rotation, VIS closes its shutter during the remaining exposures while NISP photometric imaging is performed. At the end of the last frame, a slew towards the next field is performed. A significant part of the mission is devoted to instrument calibrations and sample characterisation. The former class needs specific observations, ~6 months in total, on given targets (spectro-photometric standards, repeated fields for stability and flats). The latter needs repeated observation for depth and have different dispersion angles for the same objects or observations on well-known astronomical fields, ~6 months in total. Interleaved with calibrations, much time will be devoted to the Euclid Deep Fields (EDFs), which will be two magnitudes deeper than the wide survey and cover a minimum total area of 40 deg$^2$. Because of the need of repeated observations, long observability plays a key role here and therefore possible locations are forced to be close to the ecliptic poles.

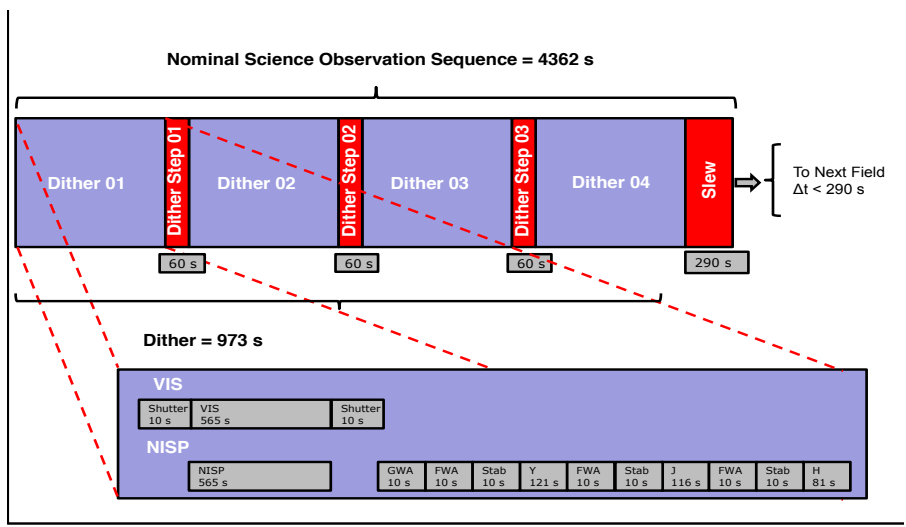

**Figure 1 Euclid standard observing sequence**

The wide area has to solve a number of demanding constraints related to the visibility: the interspersed set of calibrations, the zodiacal background (which increases by factors in going from ecliptic poles to the ecliptic equator and, moreover, is also time dependent), galactic dust extinction and scattered light. The resulting Euclid sky coverage has to exclude the ecliptic plane and the galactic plane and bulge.

Even though there are many solutions for the sky coverage, it is not obvious to find a solution satisfying all visibility, operational, and programmatic constraints. An optimal solution was found for the MPDR by the Euclid Consortium Survey group using a novel software, ECTile, developed exclusively for this purpose. The algorithm takes into account all the constraints and optimises possible sequences of pointing's on the sky. In practice the time reserved for calibrations is set as an input and the best coverage is found by using the unallocated days to observe the not yet covered sky, weighted with a merit function.

The solution, shown in Figure 2 with colour coding according to the epochs, is close to the theoretically maximum achievable (a straight line in the second plot). Towards the end of the survey most of the visible sky has been observed previously, so that no new areas have very limited visibility, causing only small increases in the growth curve.

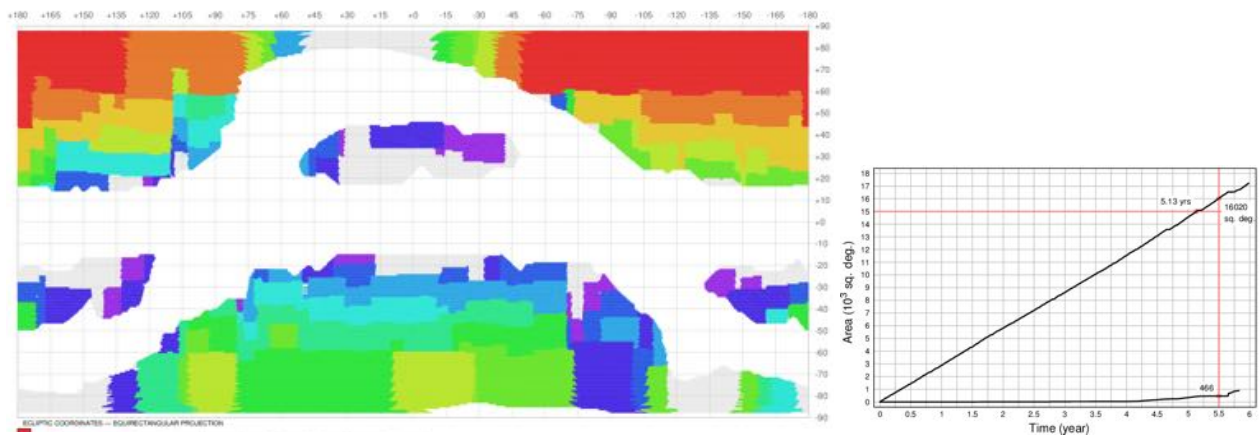

**Figure 2 Left panel: area covered by the wide survey (ecliptic coordinates, colour coding follows the epoch of observation). The empty regions reflect the ecliptic equator and the galaxy plane Right panel: growth curve, the increase of the area covered by the wide survey as a function of time.**

It must be recalled that the above reference survey is a proof of feasibility, the final survey will be delivered after launch and in-orbit performance verification.

## 3. SPACECRAFT DESIGN

The spacecraft can be subdivided in three main parts: a Service Module, a Payload Module, including the telescope, and the Scientific Instruments. They are separately described in the following sections.

### 3.1 Service Module

The Service Module (SVM) comprises the spacecraft subsystems supporting the payload operation, hosts the warm electronics of the payload, and provides structural interfaces to the Payload Module (PLM) and the launch vehicle. The Sunshield, part of the SVM, protects the PLM from illumination by the sun and supports the photovoltaic assembly supplying electrical power to the spacecraft. The overall spacecraft envelope, compatible with the Soyuz ST fairing, fits within a diameter of 3.74 m and a height of 4.8 m, see Figure 3.

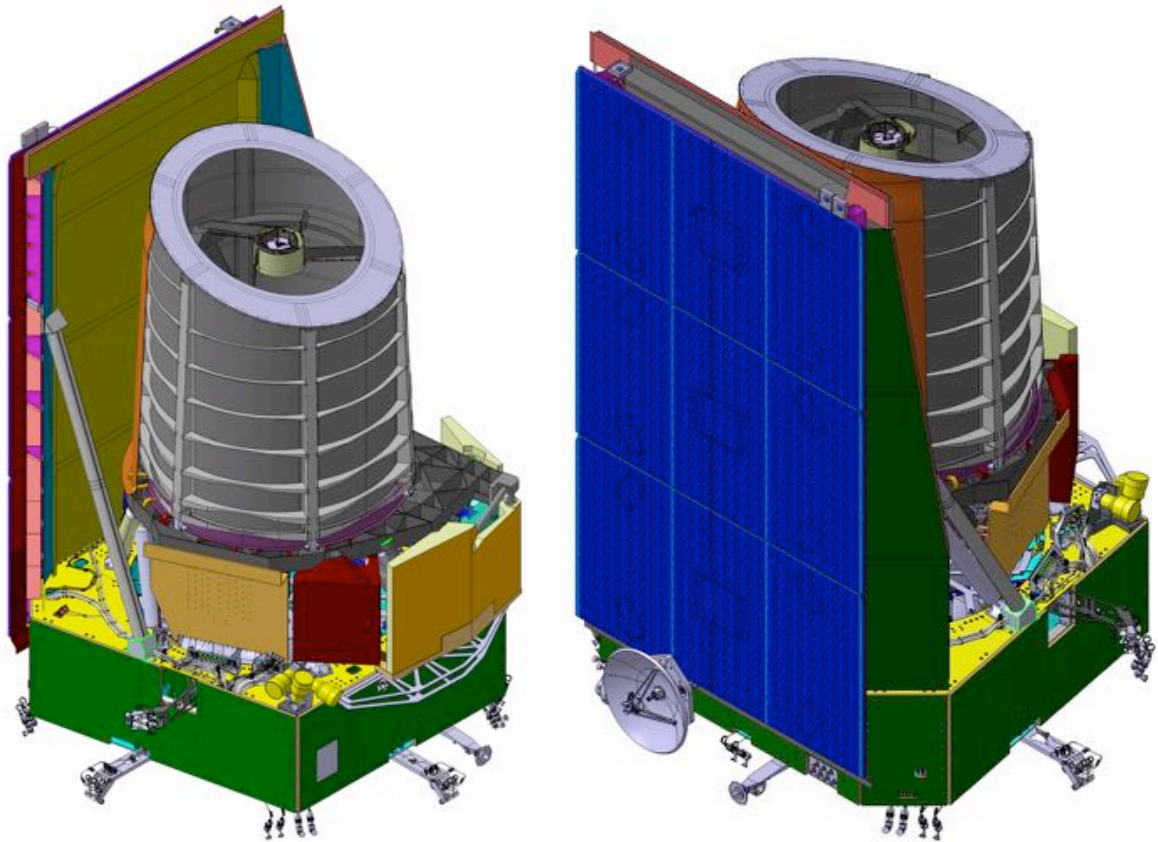

**Figure 3 Euclid spacecraft overview**

Mechanical and Thermal Architecture

The SVM (Figure 4) is an irregular hexagonal base built around a central cone that provides the interfaces with the launcher and with the PLM and encloses the Hydrazine and Cold Gas propellant tanks. External equipment attached to the SVM include a high-gain antenna, three low-gain antennas, hydrazine and cold-gas thrusters, on dedicated pods to enhance thrust efficiency, and sun sensors.

The SVM accommodates the equipment grouped on six side panels (Figure 5) according to functions: (Telemetry and Telecommand, (TT&C), Attitude and Orbit Control (AOCS), Central Data Management (CDMS) and Electric Power (EPS), payload and Fine Guidance Sensor (FGS) warm electronics). Each panel can be individually dismounted to ease the integration of equipment and their access. The SVM footprint and the sunshield are designed to keep the PLM in the shadow within the allowed range of variation of the sun direction in the spacecraft reference frame. The PLM is structurally connected to the SVM via three glass-fibre bipods attached to the SVM in six points along the central cone upper ring of 2.25 m diameter and in three points to the baseplate of the PLM. This connection forms an isostatic mount preventing the SVM induced distortion to affect the PLM structure. The lower ring of the central cone is connected to the launcher vehicle adapter interface, a standard 1666 mm adapter, and locked at launch with a low shock separation clamp band.

The Sun Shield (SSH) consists of a carbon fibre reinforced plastic (CFRP) frame made of 2 vertical poles with diagonal stiffeners and 2 struts slanting toward the SVM. The front side carries the photovoltaic assembly in 3 identical panels, while the rear side is covered by Kapton MLI. On the top edge, an optical baffle consisting of 3 blades with decreasing height has the function of attenuating the sunlight diffracted towards the PLM down to negligible levels. A dedicated feature is embedded on a corner of the SSH to provide extra radiation shielding to the VIS Instrument focal plane.

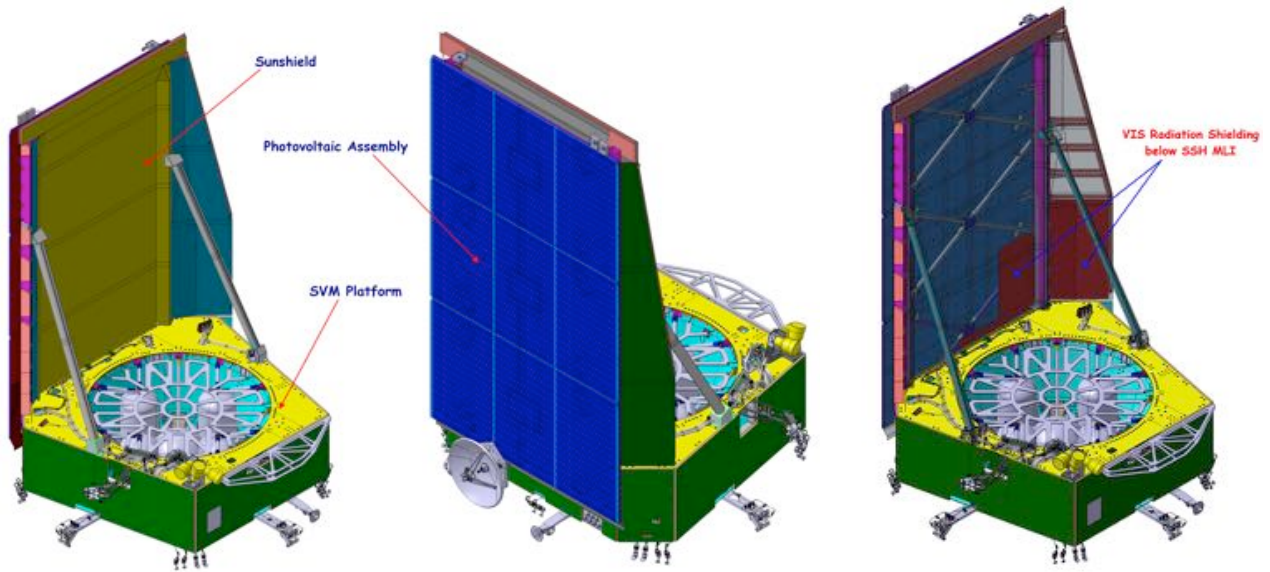

**Figure 4 SVM overview (the central cone aperture is sealed by MLI belonging to the SVM, not shown here**

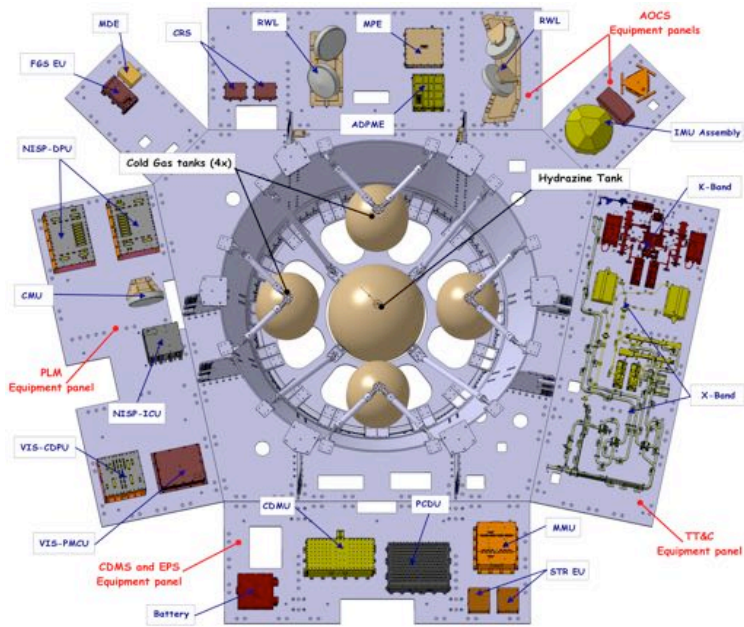

**Figure 5 SVM equipment accommodation**

The thermal control is based on a passive design using radiators, multilayer insulation (MLI) and heaters operated in Pulse Width Modulation. The design drivers are the short-term temperature stability of the PLM conductive and radiative interface under the maximum commanded Solar Aspect Angle (SAA) change, and minimal (<25 mW) heat flux into the coldest NISP radiator. High performance Kapton MLI is installed on the on SVM top floor, PLM bottom and Sun Shield rear side to minimise the heat flux and thermal disturbances onto the PLM.

Electrical and Data Handling Architecture

The spacecraft provides 28V regulated power to equipment and instruments electronics through protected lines individually commandable provided by the on-board power conditioning and distribution unit (PCDU). The PCDU also provides power to the heaters, to the pyro actuators and controls the charge and discharge of the battery. The battery is used only during the launch phase and is design to provide up to 419 W of power and a total energy of 300 Whr. In the other phases of the mission the sun shield three panels provide a power between 2430 and 1780 Watt depending from the spacecraft orientation and ageing of the panels. See budgets in Table 2.

One centralised on-board computer (Command and Data Management Unit, CDMU) provides spacecraft and AOCS command, control and data processing. The CDMU is a modular unit including standard core boards plus dedicated I/O boards to interface AOCS and spacecraft units and devices. The Processor Module is based on a general-purpose space qualified microprocessor (LEON-FT) with minimum computational power of 40 MIPS and 5 MFLOPS. Two processor modules are comprised in a single-failure tolerant unit.

The number of scientific exposures and high-resolution images generate a high science data volume and require large on board memory capable of hosting the 850 Gbit of daily generated. The on-board Mass Memory Unit (MMU) has a capacity of 4Tbit EoL sufficient to store 72 hrs of scientific data and 20 days of spacecraft housekeeping. The MMU stores instruments data and housekeeping and other ancillary data in named files organized in a two level folders' structure.

The commands and telemetries are distributed and collected mostly via two standard Mil-Std-1553 buses, one dedicated to the spacecraft equipment and another to the instruments and mass memory, although some spacecraft equipment have dedicated connections. The instruments deliver high volume scientific data via high speed SpaceWire links directly into the mass memory. The platform bus handles non-packet remote terminals (RT) and the FGS, and is characterised by cyclic communication frames at 10 Hz, linked to the AOCS control cycle. The transfer layer protocol of the science bus is based on a cyclical communication frame at 60 Hz, maximising the efficiency of data transfer per communication frame.

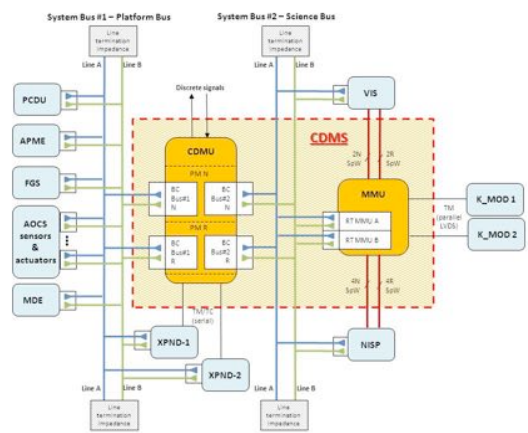

Figure 6 CDMS interfaces

Files stored in the mass memory are downloaded using the standard CCSDS File Delivery Protocol (CFDP) using the reliable transfer with acknowledges for the downlink and the simple unreliable transfer for uplink. Both the X and K band communication link can be used for the file transfer. The baseline configuration expect the directives of the CFDP to be transmitted via X-band to ensure visibility of the file downlink in progress by Ground also in case of adverse weather conditions, while data are downloaded via K-band to maximise the data rate. Any other combination is however possible.

Telecommunications

The telecommunications architecture includes two independent sections: an X-band section used for telecommands, monitoring and ranging and a K-band section dedicated to high rate telemetry (Figure 7).

The X-band section supports uplink of telecommands at two different rates (4 kbit/s and 16 kbit/s), downlink of real time housekeeping TM at two different information rates (2 kbit/s and 26 kbit/s), and standard ranging. The X-band section uses two X-band transponders, with receivers operated in hot redundancy and cross-coupled with CDMU TC decoders and transmitters operated in cold redundancy and cross-coupled with CDMU TM encoders. Three X-band LGA's with hemispherical coverage are used. Two of them (LGA-1 and -2), placed on opposite sides of the spacecraft and working in opposite circular polarisations, provide the omnidirectional coverage. LGA-3, mounted to the HGA support structure and sharing its pointing mechanism, supports high rate telecommand. The receivers are cross-coupled with LGA-1 or LGA-3, and LGA-2.

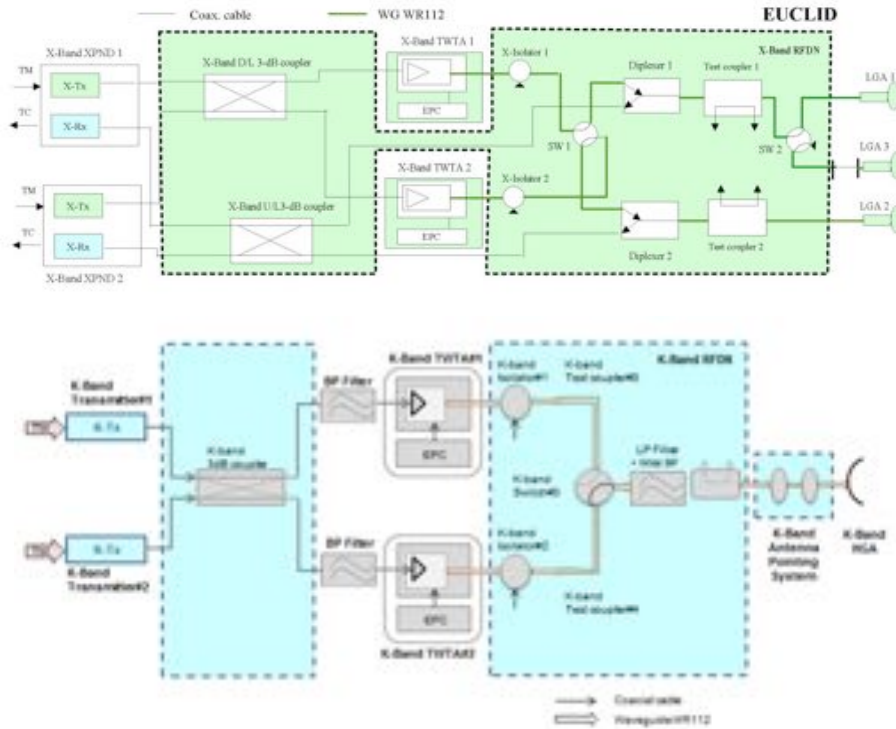

Figure 7 TT&C block diagrams, X-band (top) and K-band (bottom)

The K-Band section supports downlink of recorded science and housekeeping telemetry at two different data rates: nominal at 73.85 Mbit/s and reduced at 36.92 Mbit/s to allow 3 dB extra link budget margin in adverse weather. The K-Band section features two K-band modulators operated in cold redundancy, directly connected with the mass memory unit for the download of files. See RF link budgets in Table 3. A 70 cm diameter K-band High Gain Antenna (HGA) is installed on a two-degree-of-freedom pointing mechanism with angular range of ±55° in azimuth and -70° to + 40° in elevation.

Attitude and Orbit Control

The Euclid image quality requirements demand very precise pointing and small jitter, while the survey requirements call for fast and accurate slews. The attitude and orbit control system (AOCS) provides 75 mas (milli-arcseconds) Relative Pointing Error (RPE) over 700 sec and 7.5 arcsec of Absolute Pointing Error both with 99.7% Confidence Level. See detailed pointing performance in Table 4. A Fine Guidance Sensor (FGS) with 4 CCD sensors co-located within the VIS imager focal plane provides the fine attitude measurement. Cold gas thrusters with micro-newton resolution provide the forces used to actuate the fine pointing. The Gyro and FGS-based attitude control rejects the low frequency noise (less than 0.1 Hz) ensuring that the RPE requirement is met. Three star trackers (STR) provide the inertial attitude accuracy to comply with the APE requirement. The STRs are mounted on the SVM and thus subject to thermoelastic deformation when large slews are executed. To solve this problem, the FGS will be endowed with absolute pointing capabilities (based on a reference star catalogue), allowing cross-calibration of STR and FGS reference frames. A high performance gyroscope is included to reduce high frequency attitude estimation noise, to manage FGS delays and recovery from temporary outages. Four reaction wheels execute all the slews (field slews, 50-100 arcsec dithers, and large slews between different sky zones). After each slew manoeuvre the wheels are controlled to slow down until friction stops them. Keeping the reaction wheels at rest during operation ensures noise-free science exposures by eliminating the micro-vibration associated to reaction wheel actuation.

The micro-propulsion employed for fine attitude control is based on cold-gas Nitrogen thrusters in a configuration of two branches with six thrusters each. Four high-pressure tanks provide storage of 70 kg Nitrogen, sufficient for 7 years operation with nearly 100% margin.

Orbit control and attitude control in non-science modes are actuated by two redundant branches of ten 20N hydrazine thrusters. In each branch, two thrusters, one on either side of the spacecraft, provide torque-free thrust for the Trajectory Control Manoeuvres on the way to SEL2, monthly Station Keeping Manoeuvres at SEL2, and disposal at end of life. The other eight thrusters provide force-free torques for angular momentum and attitude control in non-science modes. Hydrazine storage is provided by one central tank with 137.5 kg propellant mass capacity with 10% volume margin over and above the prescribed delta-v margins.

System budgets

The mass budget in Table 1 shows the breakdown at module level based on detailed estimates validated by the subsystem suppliers. A lightweight adapter of the same design used on Gaia is employed.

**Table 1 System mass budget: the SVM and the PLM masses include the instrument units located in each module, reported below each item**

|  |  | Current mass [kg] |
|---|---|---|
| SVM |  | 920.6 |
| Payload warm units | 98.0 |  |
| PLM |  | 848.3 |
| Payload cold units | 193.0 |  |
| Dry mass reserve |  | 113.9 |
| Propellant |  | 199.2 |
| Launch vehicle adapter |  | 78.0 |
| Total launch mass |  | 2160.0 |

**Table 2 System power budget. The sizing case is shown (maximum payload power demand, communications on; end of life; maximum voltage).**

|  | Power [W] |
|---|---|
| SVM (@ 28 VDC) | 774 |
| PLM (@ 28 VDC) | 88 |
| Instruments (@ 28 VDC) | 392 |
| System losses (3%) | 38 |
| System margin (20%) | 258 |
| Array power demand (at PCDU input and @ 29.5 VDC) | 1360 |

Table 3 RF link budget. Cebreros ground station, 1.77 million km, elevation above horizon 5° (X band) and 20° (K band). For the LGA items, degrees in parentheses indicate angle from centre of pattern.

|  | Antenna | Bit rate [kbit/s] | Nom. Margin [dB] |
|---|---|---|---|
| X band |  |  |  |
| TC (90°) | LGA1/2 | 4.0 | 4.1 |
| TC (5°) | LGA3 | 16.0 | 7.3 |
| TM (90°) | LGA1/2 | 2.0 | 5.8 |
| TM (5°) | LGA3 | 26.0 | 8.7 |
| K band |  |  |  |
| TM (low rate) | HGA | 73.85 | 4.8 |
| TM (high rate) | HGA | 36.93 | 7.6 |

Table 4 Pointing budgets (99.7% confidence level)

|  | Requirement [arcsec] | Performance [arcsec] |
|---|---|---|
| APE (X/Y axis) | 7.5 | 6.25 |
| APE (Z axis) | 22.5 | 12.27 |
| RPE (X/Y axis) | $75 \cdot 10^{-3}$ | $70.5 \cdot 10^{-3}$ |
| RPE (Z axis) | 1.5 | 0.275 |

### 3.2 Payload Module

The Euclid PLM is designed around a three mirrors anastigmatic Korsch Silicon Carbide (SiC) telescope feeding the two instruments, VIS and NISP, see schematic in Figure 8. The light separation between the two instruments is performed by a dichroic plate located at exit pupil of the telescope. The PLM is in charge of providing mechanical and thermal interfaces to the instruments (radiating areas and heating lines). Whereas NISP is a stand-alone instrument with interface bipods, VIS is delivered in several separate parts: a focal plane assembly (FPA) connected to proximity electronics, readout shutter unit and calibration unit, with dedicated mechanical and thermal interfaces with PLM.

The secondary mirror (M2) is mounted on a mechanism (M2M) for 3-DOF adjustment to compensate for launch and cool-down effects. In addition, the PLM hosts the FGS, used as pointing reference by the AOCS. All these detectors are mounted on the structure carrying the VIS focal plane, in order to ensure precise co-alignment.

Except proximity electronics of the focal planes and FGS, all electronics are placed on the SVM to minimise thermal disturbances to the PLM.

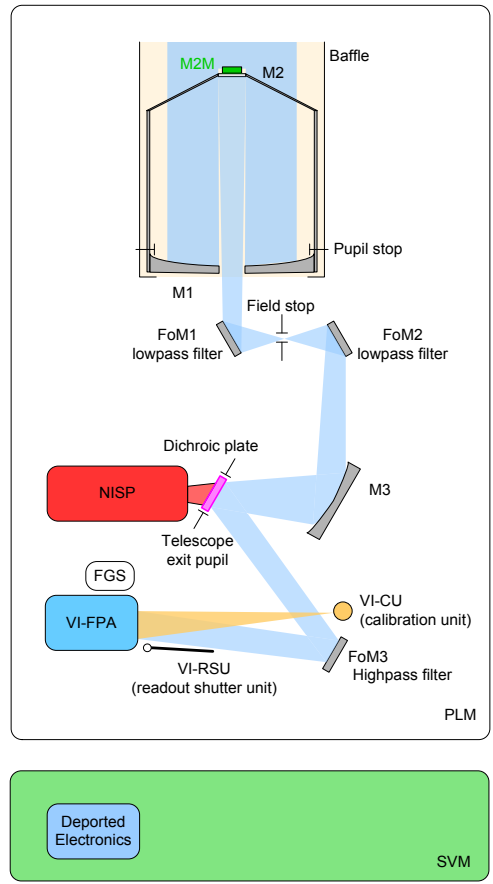

**Figure 8 Schematic functional view of the Euclid PLM**

Overall PLM architecture

The PLM is divided in two cavities, separated by the base plate:

- The front cavity including the telescope primary and secondary mirrors as well as the M2 refocusing mechanism and the associated support structure. This cavity is thermally insulated by a baffle.

- The instrument cavity including the telescope folding mirrors, the tertiary mirror, the dichroic splitter, the FGS, the two instruments (VIS and NISP), the shutter and calibration source for the VIS channel.

The PLM mechanical architecture is based on a common SiC baseplate which supports on one side telescope M1 and M2 mirrors and on other side the other optics and the two instruments. This architecture is well adapted to the selected thermal architecture with telescope and instrument cavities both passively controlled at neighbour temperatures.

The optical accommodation on the baseplate consists in implementing two folding mirrors (FoM1 and FoM2) at the entrance of the instrument cavity (between M2 and M3) to fold the optical beam in the plane of the baseplate. A third folding mirror (FoM3) allows having the VIS instrument close to an efficient radiative area.

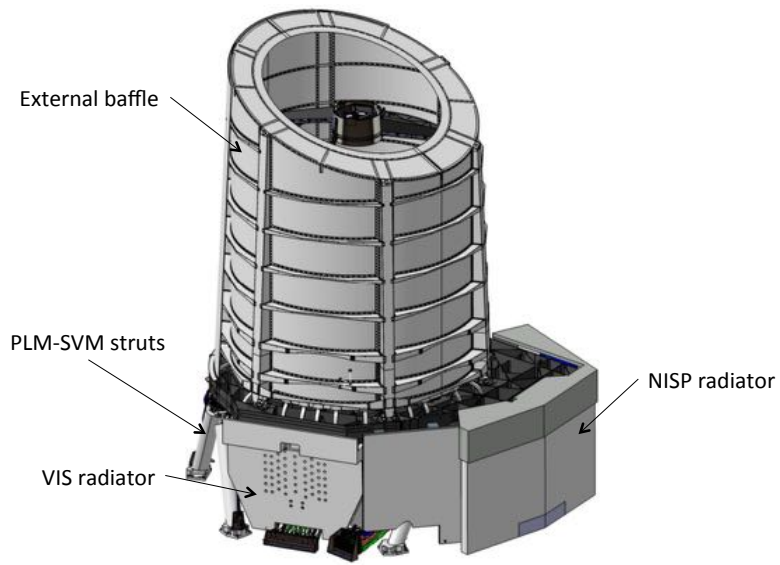

**Figure 9 External front view of the PLM**

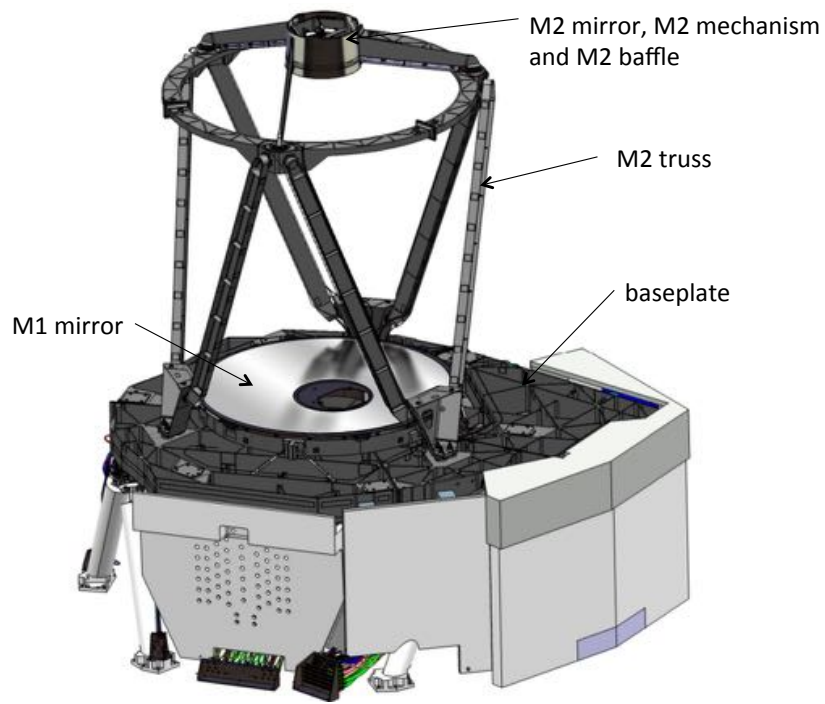

**Figure 10 External front view of the PLM without the external baffle**

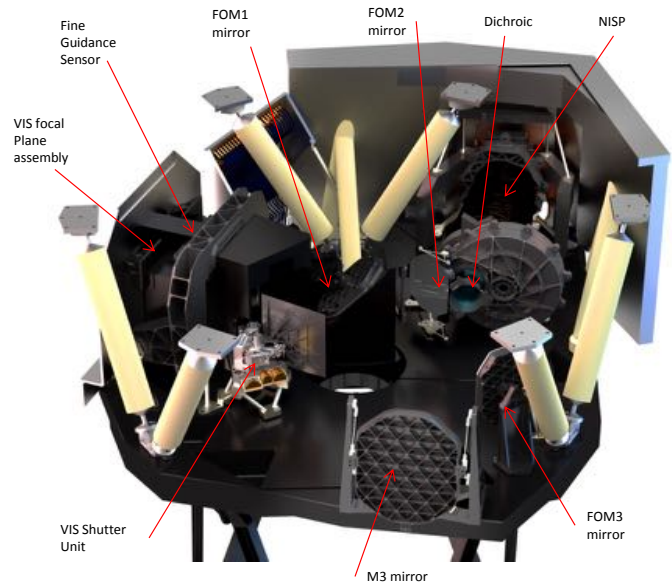

**Figure 11 Rear view of the PLM showing the instrument cavity**

PLM Optical design

The optical combination is a Korsch telescope design with a FoV offset of 0.47°. The useful pupil diameter is 1.2 m and the focal length is 24.5 m, i.e. plate scale of 8.3 mas/µm. The central obscuration has been minimised by designing a thin spider and careful design of M1 and M2 baffles. The resulting collecting area is larger than 1 m².

The challenging spectral transmission and the VIS out-of-band rejection requirements are met by complementing the dichroic plate with highpass filter on FoM1 and FoM2 and bandpass filter on FoM3.

**Table 5 Telescope fact sheet**

| Telescope type | Korsch |
|---|---|
| Focal length | 24.5 m |
| Entrance Pupil | Ø1200 mm |
| Paraxial F-Number | F/20.42 |
| M1 | Ø1250 mm |
| M2 | Ø350 mm |
| M3 | 535 x 406 mm² |
| FoM1 | 358 x 215 mm² |
| FoM2 | 283 x 229 mm² |
| FoM3 | 358 x 215 mm² |
| Dichroic plate | Ø117 mm |
| M1-M2 distance | 1756 mm |
| Useful collecting area | 1.006 m² |

| Offset along Y axis | 0.47° |
|---|---|
| VIS FoV | XAN 0.787° x YAN 0.700° |
| NISP FoV | XAN 0.779° x YAN 0.727° |
| FGS FoV envelopes | 0.116°x 0.255° |

PLM Thermal design

The passive thermal concept is adapted from the one successfully designed for Gaia PLM [5], requiring minimum heating power and providing best thermal stability. The telescope is cooled-down to its equilibrium temperature around 130K. In nominal operations, only local heating capacity at constant power is needed to adjust instruments interface temperatures to the prescribed value. The required heating power in operational mode is therefore very low, ~140 W. Additional heating lines are installed for optics decontamination during commissioning phase and for survival mode, to keep instruments in their non-operational temperature range.

This cold telescope offers high thermo-elastic stability (SiC CTE is reduced to 0.4 µm/m/K) and cold environment (135 K) to the instruments. The heat leaks to cold instruments units are minimised (actually the PLM becomes a thermal sink for all units except NISP detector and VIS-FPA electronics which are connected to out-looking radiators). Specific harness design and highly decoupled conductive and radiative thermal interfaces allow minimising the heat leaks from the SVM and the sunshield.

Image quality performances

Thanks to the high thermal and dimensional stability of the telescope, and to the capacity for accurate in-orbit alignment, the Euclid PLM design offers excellent image quality performances, with large margins on the visible Point Spread Function (PSF) ellipticity, Full Width at Half Maximum (FWHM) and near Infrared Encircled Energy (EE) radius as shown in Table 6 and Table 7. See [12] for definition of the weak lensing metric specific $R^2$ and ellipticity. This allows improved mission performances and offers flexibility for relaxed image motion requirements, i.e. reduced pointing stability and larger tolerance to micro vibrations.

Table 6 VIS image quality performance

|  | Requirement | Performance |
|---|---|---|
| Ellipticity | < 0.11 | 0.081 |
| FWHM | < 137 mas | 135.4 mas |
| $R^2$ | < 0.053 as² | 0.051 as² |

Table 7 NISP image quality performance

|  | Wavelength [nm] | Requirement [µm] | Performance [µm] |
|---|---|---|---|
| rEE50 : |  |  |  |
|  | 2000 | < 31.5 | 27 |
|  | 1486 | < 25.4 | 21 |
|  | 1033 | < 19.2 | 16.8 |
| rEE80 : |  |  |  |
|  | 2000 | < 77.0 | 74.9 |

| | | |
|---|---|---|
| 1486 | < 58.0 | 56.7 |
| 1033 | < 43.0 | 41.8 |

### 3.3 Scientific Instruments

**The VIS instrument**

The VIS (visible imager) instrument is optimised to register resolved images of galaxies in the 550-900 nm passband. See [6] for more details. The VIS nominal survey images have a 10σ extended source detection limit of AB 24.5 mag and are used to determine the shapes of at least 30 galaxies per arcmin$^2$ over the survey area.

The VIS instrument consists of a CCD [7] based focal plane array (FPA) employing one wide visible passband, a shutter mechanism to close the optical path for read out and dark calibration, and a calibration unit for flat field measurements. See FPA location in the PLM in Figure 11. The FPA supports 6×6 CCDs (4k×4k pixels each) with 0.10 arcsec pixel plate scale, giving a geometric field of greater than 0.5 deg$^2$ including the gaps between the detectors. The VIS data processing and instrument control units are mounted in the spacecraft service module. The VIS central data processing unit collects data from 144 CCD quadrants, arranges all the pixels in the correct order and compresses this very large image (24k x 24k), in approximately 250 seconds. The VIS power and mechanisms control unit activates the shutter and the calibration unit. To have full control over the sources of systematic errors no additional VIS image processing will be done on board, all CCD data is transferred to ground.

**The NISP instrument**

The NISP instrument (see [8] for more details) is designed to carry out slitless spectroscopy and imaging photometry in the near-infrared (NIR) wavelength. The NISP spectroscopy supports the galaxy clustering probe and is optimised to measure the redshifted Hα (rest wavelength 656.3 nm) emission line of galaxies with redshifts between *z=0.8* and *z=1.8*. NISP will detect redshifts of at least 1700 galaxies/deg$^2$ in the corresponding wavelength range 1250-1850 nm, at a detection limit of $2\times10^{-16}$ erg/s/cm$^2$ (4.5σ) for a typical source of 0.5 arcsec. The imaging photometry supports the weak lensing probe by photometric measurements of galaxies down to AB 24 mag (5σ point source) in three Euclid bands: Y, J, H. See NISP location in the PLM in Figure 11.

NISP contains an array of 4×4 HgCdTe NIR detectors (2k×2k pixels each) with 0.3 arcsec per pixel and 2.3 μm cut-off wavelength. The sensor chip system, consisting of the sensor chip assembly, an ASIC read-out electronics, and flex cable connecting each sensor to each read-out electronic unit, is provided to Euclid by NASA/JPL and is manufactured by Teledyne Imaging Systems (Camarillo, CA).

NISP can be operated in either photometer mode or slitless spectrometer mode by means of a filter wheel and a separate grism wheel. In the slitless spectrometry mode, the light is dispersed by grisms covering the wavelength range 0.92 to 1.85 μm with a constant Δλ at a spectral resolution λ/Δλ > 380 for an object of 0.5 arcsec diameter. NISP will carry out the detector voltage ramp processing on board providing one photometric or spectroscopic image per exposure. This sets a high demand on the processing capabilities and the power consumption of the NISP warm units.

## 4. THE EUCLID GROUND SEGMENT

The Euclid ground segment consists of two blocks: the Operational Ground Segment (OGS), comprising the Mission Operations Centre (MOC) and the grounds station network, managed entirely by ESA, and the Science Ground Segment (SGS), for which the management and the implementation is shared between ESA and the Euclid Consortium. The Science Ground Segment (SGS) is responsible for the data processing, instrument operations, survey definition and archiving. The MOC is responsible for the overall mission operations. One Ground Station supports a daily telemetry communications period of nominally 4 hours during nominal operations and longer during the commissioning and performance verification phases. Euclid has X and K band transponders to support the tele-commanding and the science data transfer to ground, respectively. The K band section supports a downlink data volume of 850 Gbit of compressed science data in the allocated 4 hours.

Euclid will deliver an unprecedented large volume of data for astronomical space missions: more than 1 Pbit of data per year, about 4 times more data than Gaia. Furthermore, a large volume of ground-based data from optical surveys like DES, LSST or others is used for calibrations, quality control tasks and scientific data reduction, specifically for obtaining photometric redshifts. The ground based data have to undergo Euclid specific processing (such as the conversion to common astrometric and magnitude reference systems) in order to be consistently handled with the Euclid data.

## 4.1 The Operational ground Segment

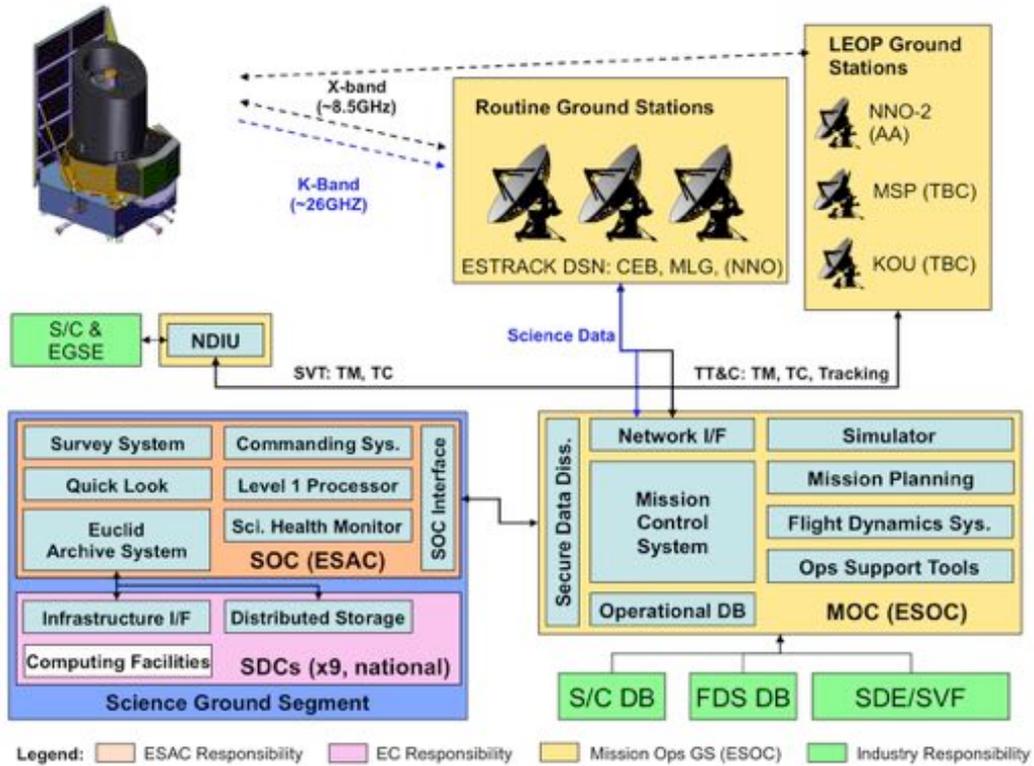

**Figure 12 Euclid ground segment overview**

The Euclid ground segment comprises the Mission Operations Ground Segment (OGS) and Science Ground Segment (SGS). Its main components are shown in Figure 12:
1) The Mission Operations Centre (MOC), located in ESOC, Darmstadt, in charge of all mission operations planning, execution, monitoring and control of the spacecraft and operations ground segment.
2) The Ground Stations, which belong to ESA's tracking network, are under control of the MOC at ESOC. The network is composed of 3 Deep Space antennas in New Norcia (NNO), Western Australia, Cebreros (CEB), Spain, and Malargüe, Argentina, which are used during Launch and Early Operations Phase (LEOP), the commissioning phase and the routine mission. In addition, a small X-band antenna is available in New Norcia (NNO) for first acquisition during LEOP. If still available at the time some of the 15m ESTRACK stations in Kourou (KOU), French Guiana, and Maspalomas (MSP), may also be used during LEOP. The ground stations provide the capability to communicate with the spacecraft during all mission phases post-launch in X-band on up- and downlink. Two of the 35m stations will also be upgraded for reception of Euclid's 26GHz high data rate signal. The stations will also be used to perform ranging and Doppler measurements with the spacecraft for orbit determination purposes.
3) The Science Operations Centre (SOC), located in ESAC, Villafranca, is in charge of scientific operations planning, performance monitoring of the payload using spacecraft and instrument files delivered by the MOC, and interface with the Euclid Science Data Centres (SDC), science data archiving and distribution and scientific analysis support (see section 4.2).
4) The Communications Network, linking the various remotely located centres and stations to support the operational data traffic.
5) The industrial prime contractor who provides the spacecraft and flight dynamics databases as well as onboard software images to the MOC.

There are a number of Euclid specific operational challenges that are driving the design of the Euclid ground segment and its operations:

- A mission which will fly a large amplitude quasi-halo orbit around the SEL2;

- Very high volume and data rates (~850Gbits/day) to be downlinked during 4 hours long ground station contacts per day;

- Downlink in 26GHz (K-band) for large data volumes, which is susceptible to relatively high atmospheric attenuations in case of bad weather, and 8GHz (X-band) for real-time TM, TC and ranging/Doppler;

- Implementation of the CCSDS CFDP protocol for downlink of files (science and on-board recorded Housekeeping TM) and file system in the on-board Mass Memory Unit. The implementation of the protocol on-board requires a close hand-shake with the ground segment;

- An autonomous spacecraft with very stringent pointing requirements and a complex AOCS including a dedicated Fine Guidance Sensor (FGS), which is using a custom-built star catalogue.

**4.2 The Science Ground Segment**

The initial concepts for the Science Ground Segment have been described in [9]. ESA provides the SOC, run by ESAC, in charge of survey planning, first consistency and quality checks, and delivery of data for public use. SOC is the only interface to MOC: the latter provides raw telemetry and all auxiliary information necessary to manage the mission from the scientific point of view; vice-versa, the SOC provides the MOC with information related to observation planning and instruments commanding.

The Euclid Consortium (EC) provides the fraction of the Ground Segment (the Euclid Consortium Science Ground Segment - ECSGS) performing the data processing from telemetry down to the mission data products. It is physically composed of a number of Science Data Centres (SDCs), in charge of instrument-related processing, production of science data products, simulations, ingestion of external data and in general all science-driven data processing. Two Instrument Operation Teams (IOTs), one for each instrument, guarantee instrument maintenance and operations: coordination of the IOTs between them is guaranteed, so to allow to have a single interface between the ECSGS and the SOC. The computational needs of the IOTs are supported by the SDCs.

In Figure 12, the box on the lower-left part of the graph representing the Science Ground Segment can be functionally expanded in a set of more detailed functions. The Euclid Archive is at the heart of the ground segment, since it is designed to provide centralised data management function all the elements of the SGS and the science archive for both the EC and general scientific community.

The processing of science data can be decomposed in ten logical data Processing Functions, defined by considering that they represent self-contained processing unit (i.e. they represent the highest-level breakdown of the complete pipeline that can be achieved with units that communicate only with the help of the archive. The identified Processing Functions are listed in the following, and their interrelations are shown in Figure 13.

- LE1, in charge of telemetry unscrambling;

- VIS, in charge for processing the Visible imaging data from edited telemetry to fully calibrated images, as well as source lists (for quality check purposes only);

- NIR, in charge of processing the Near-Infrared imaging data from edited telemetry to fully calibrated images as well as source lists (for quality check purposes and to allow spectra extraction);

- SIR, in charge of processing the Near-Infrared dispersed imaging data from edited telemetry to fully calibrated spectral images and extracted spectra;

- EXT, in charge of entering in the Euclid Archive all of the external data that are needed to proceed with the Euclid science;

- SIM, providing the simulations needed to test, validate and qualify the whole pipeline;

- MER, performing the merging of all information, providing stacked images and source catalogues where all the multi-wavelength data (photometric and spectroscopic) are aggregated;

- SPE, extracting spectroscopic redshifts from the SPE spectra;
- PHZ, computing photometric redshifts from the multi-wavelength imaging data;
- SHE computes shape measurements on the visible imaging data;
- LE3 is in charge of computing all the high-level science data products (Level 3), from the fully processed shape and redshift measurements (and any other possibly needed Euclid data).

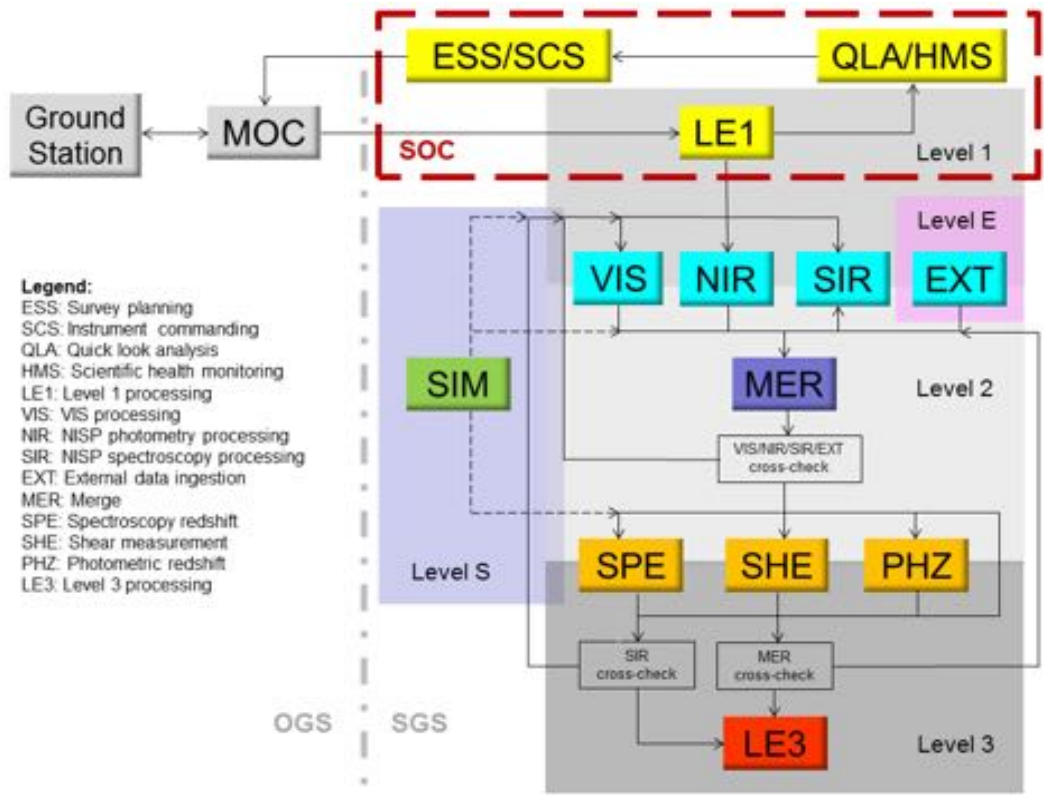

**Figure 13 Functional decomposition of the Euclid Science Ground Segment**

Each of the Processing Functions is supported by an Organisation Unit that, starting from the requirements flowed down from the higher levels and placed on the specific Processing Function, perform algorithmic research by designing prototypes, performing numerical tests, and comparing the results with the original requirements. Once validated by the proper OU, the prototype is passed on to an SDC, along with a test harness, and the prototype is turned into a full-fledged Euclid pipeline element.

From the above description, it is clear that the SGS activity is geographically spread as part of its very design. It is also to be noted that the SOC and the ECSGS are designing and developing a single and integrated SGS. As part of the joint work being performed by staff from ESA (SOC) and the EC (ECSGS) to build the Euclid SGS, a SGS System Team (SGS ST) common to both SOC and ECSGS has been active for several years, taking the lead in helping the SGS to define the overall data processing philosophy, architecture and strategy. Among the tasks of the SGS ST: preparing coding guidelines; defining and implementing tools to support software tests and integration; defining, designing, implementing and testing common software (e.g. interfaces, transfer systems and common toolboxes); defining, designing, implementing and testing the archive; designing and implementing tools to define and maintain an Euclid common Data Model.

The soundness of this development schema is periodically checked. Not only thorough formal reviews and Technical checkpoints, but also through the definition and running of ICT and Scientific (data processing) Challenges. This mechanism is aimed at verifying practically that the system being designed and incrementally implemented is sound, technically feasible, scalable and capable of being run in a distributed environment providing consisting results. Challenges are implemented at all SDCs, and are not considered as passed until all SDCs fulfil the requirements. Five ICT Challenges and the first Science Challenge have been successfully completed, while the sixth ICT Challenge and second Science Challenge are underway at the time of writing.

## 5. SUMMARY OF MISSION PERFORMANCE

The Euclid mission science objectives translate into stringent performance requirements allocated at the different elements of the mission architecture. A number of mission level mathematical models and simulations were used during the initial phases leading to the mission SRR in December of 2013, to derive the full set of requirements for the instruments (VIS and NISP), the PLM, the spacecraft pointing and stability, the data processing algorithms, the survey and the system calibrations. The evolution of the different elements performance is tracked against the applicable requirements, which ensures ultimately compliance to the mission needs. However, to evaluate the current performance of the system, the estimates of the individual elements have to be combined at mission level. This assessment has been performed in the framework of the Mission PDR in mid 2015.

The Euclid mission level performance main requirements can be broadly grouped in the following categories:
- *Image quality:* requirements on the complete VIS and NISP observing chain Point Spread Function (PSF) characteristics;
- *Radiometric performance:* expressed as photometric sensitivity achievable for the visible and near infrared channels and as line detection flux limit for the spectrometric channel;
- *Spectroscopic performance:* defined for Euclid in terms of purity (proportion of accurate redshift determination) and completeness (ratio of successfully measured targets from the observable galaxy region).

Below, the current observed performance for each requirement group is presented.

**Image Quality (IQ)**

The IQ requirements for Euclid are defined in two different ways for the visible and infrared channels. The infrared spectrophotometer performance is expressed in traditional terms of *Encircled Energy* (EE) since the main purpose is photometric light detection (either in a band or of a spectral feature). The visible channel however is devoted to the measurement of weak lensing by means of observation of the shear of galaxies. This imposes constraints on the *morphology* of the optical response of the system (see [1] for details). To achieve that, the PSF shape has to be controlled and requirements are expressed in terms of FWHM, maximum ellipticity ($\varepsilon$) and $R^2$ (related to the energy in secondary lobes of the PSF).

For both channels, the PSF quality depends on: the telescope performance (as described in section 3.2), the AOCS system residual pointing jitter during an observation, the instrument optics quality and the detection element response. For the VIS channel, the complete system PSF can be expressed as:

$$PSF_{system}(\theta_X, \theta_Y, \lambda, I) = PSF_{telescope}(\theta_X, \theta_Y, \lambda) * PSF_{instr}(\theta_X, \theta_Y, \lambda) * PSF_{aocs}(\theta_X, \theta_Y) * PSF_{detector}(\lambda, I)$$

The optical response of the system is in general both wavelength and field point dependent. In the case of the Euclid CCDs of VIS, the response has shown to be dependent on the intensity of the observed energy [10].
At this stage of the development, the design of the optical system is completed, however the alignment and manufacturing tolerances have to be taken into consideration. That is also the case for the design of the AOCS system. For that purpose, the calculation of the system performance is based on a Monte Carlo analysis combining: realizations of the optical system at several points in the FoV, 100 realisations of pointing jitter series over a nominal exposure, and the best fit CCD PSF determined by Bayesian analysis in [10]. The process is summarized in Figure 14, and a sample realization of the VIS system PSF is displayed in Figure 15. A similar process was followed for the NISP channel.

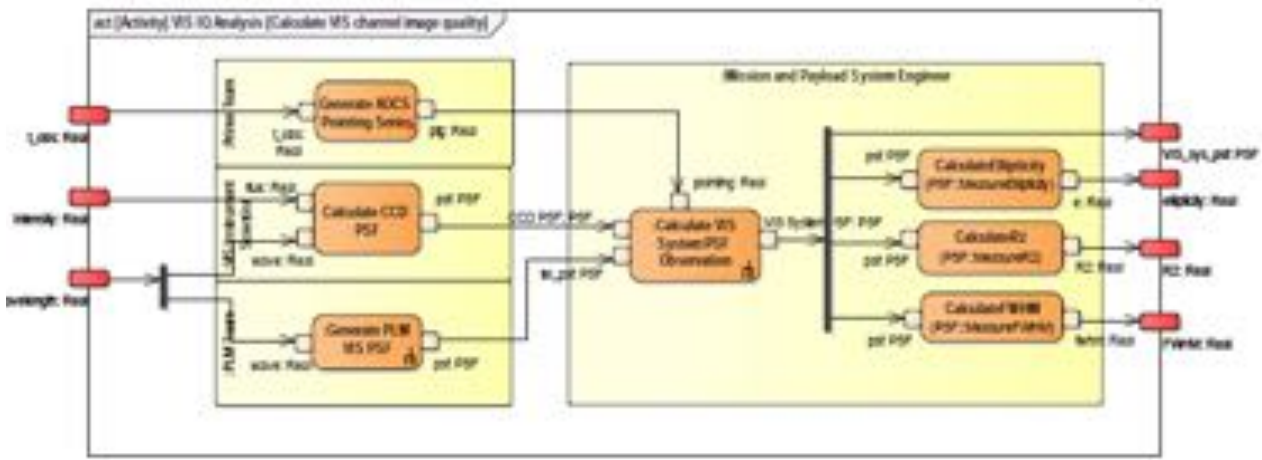

**Figure 14 - VIS Image quality derivation**

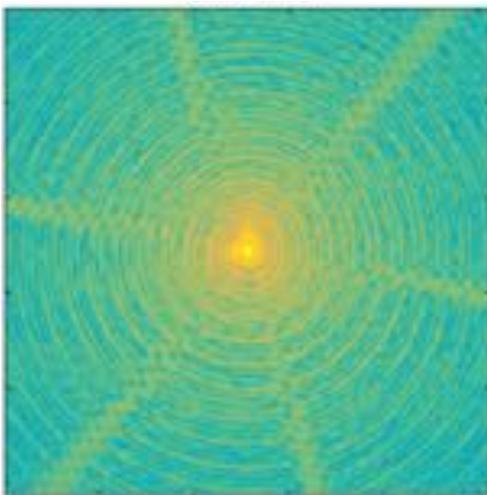

**Figure 15 - VIS channel system PSF realization (log display)**

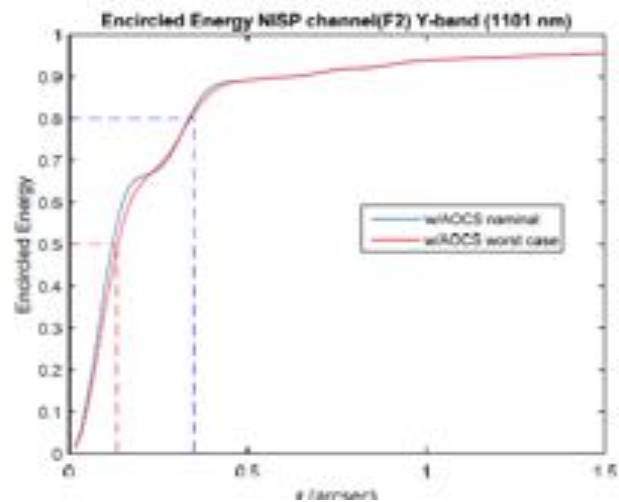

**Figure 16 - NISP photometric Y-band channel system encircled energy (for field point F2 - center of field)**

With the calculated PSFs, the performance parameters are calculated at a 3σ level for a discrete number of field points. Results at Mission PDR are shown in Table 8. The requirements are met with variable margins for the different parameters in both VIS and NISP channels. A couple of points are worth noting: the ellipticity performance of the system is better than the telescope stand-alone values (compare with Table 6). This is due to the introduction of the AOCS nearly gaussian jitter, which circularizes the PSF removing the local variations from the pure optical PSF. Another is the requirements on ellipticity stability marked in yellow. The requirements depend on the thermo-elastic performance of the telescope and the spacecraft and the level of correction that can be achieved on ground data processing. Current estimations show compliance, but final confirmation will be provided in the complete structural-thermo-optical (STOP) analysis to be completed by Mission CDR and the initial assessment of ground segment PSF determination algorithms under development.

**Table 8 Euclid mission level image quality performance at Mission PDR**

| Technical Performance Measure | | Requirement | CBE |
|---|---|---|---|
| **Image Quality** | | | |
| VIS Channel | FWHM (@ 800nm) | 180 mas | 163 mas |
| | ellipticity | 15.0% | 5.9% |
| | R2 (@ 800 nm) | 0.0576 | 0.0530 |
| | ellipticity stability $\sigma(\epsilon_i)$ | 2.00E-04 | 2.00E-04 |
| | R2 stability $\sigma(R2)/<R2>$ | 1.00E-03 | 1.00E-03 |
| | Plate scale | 0.10 " | 0.10 " |
| | Out-of-band avg red side | 1.00E-03 | 1.13E-05 |
| | Out-of-band avg blue side | 1.00E-03 | 2.12E-04 |
| | Slope red side | 35 nm | 15 nm |
| | Slope blue side | 25 nm | 8 nm |
| NISP Channel | rEE50 (@1486nm) | 400 mas | 217 mas |
| | rEE80 (@1486nm) | 700 mas | 583 mas |
| | Plate scale | 0.30 " | 0.30 " |

**Radiometric Performance**

The sensitivity limit achievable in the Euclid channels is limited by the background noise, with two major contributors: the Zodiacal background level and the diffused straylight both in-field and out-of-field reaching the focal plane of the instruments. Both these components are variable across the portion of the sky being imaged and also change seasonally. The straylight induced degradation of the signal-to-noise ratio and light-pollution of spectrum between two close objects in the sky has a direct impact on the number of valid observed objects in the survey. For a sensitive wide field telescope as Euclid, even the sources situated out of the field of view can contribute to a great amount through indirect scattering in non-optical surfaces. Significant attention and effort at system level has been dedicated to define a correct approach for end-to-end straylight levels assessment and control (see more information in [11]).

At mission level, we must therefore combine the selected survey strategy and pointing maps (Figure 2), the system straylight response, the sky sources contributing to the background (from the Milky Way and external) and the Zodiacal light to determine the total background levels seen by the instruments. This process is summarised in Figure 17.

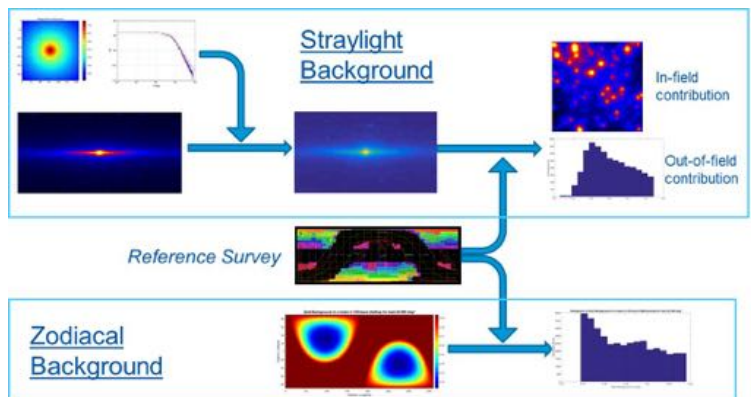

**Figure 17- Background radiation determination process**

Signal levels are determined by the systems transmission and detection elements quantum efficiency, as well as by the selected radiometric aperture for signal recovery. The selection of radiometric aperture however needs to be balanced with the background: a recovery over a larger pixel area increases the background noise, reducing the effective Signal to Noise Ratio.

Ultimately, based on the signal and background levels, SNR maps over the complete observable sky survey have been produced to confirm that 15,000 deg$^2$ can be observed with the required SNR (see samples in Figure 18 and Figure 19). Table 9 summarises the Signal to Noise ratio for the different Euclid channels over the best 15,000 deg$^2$ observable areas of the sky for each of the cases at the Mission PDR analysis. All show good margins with respect to the requirement at this stage.

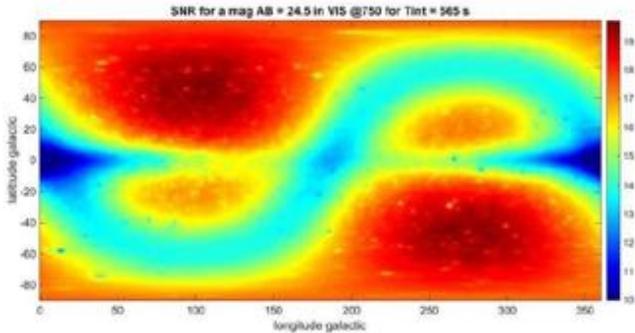

Figure 18 - SNR map for VIS Channel (for a source of mAB=24.5)

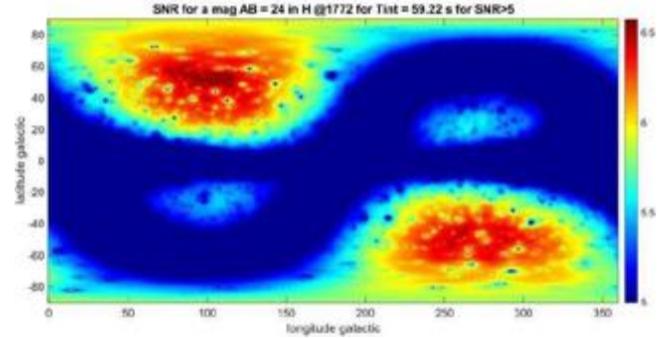

Figure 19 - SNR map for NISP photometric H channel (for a source of mAB=24)

Table 9 Euclid system sensitivity performance at Mission PDR

| Sensitivity | | | |
|---|---|---|---|
| VIS SNR (for mAB = 24.5 sources) | | 10 | 17.1 |
| NISP-S SNR (@ 1.6um for 2xe-16 erg cm-2 s-1 source) | | 3.5 | 4.87 |
| NISP- P SNR (for mAB = 24 sources) | Y-band | 5 | 5.78 |
| | J-band | 5 | 6.69 |
| | H-band | 5 | 5.35 |

**Spectroscopic performance**

Finally, at Mission PDR the performance of the spectroscopic channel in terms of purity and completeness achievable was evaluated based on the current spectra extraction algorithms and simulations. Slitless spectroscopy by its very nature provides a spectrum for every object within the field-of-view. This introduces confusion and additional background on the signal when compared with a traditional slit spectrometer. Purity of the sample is the fraction $p$ of measured redshifts that are correct ($N_{correct}$) out of those we can measure ($N_{meas}$). The completeness is an estimation of the number of galaxies for which we are able to measure a redshift out of the total number of object ($N_{tot}$) that can be observed at the flux limit specified.

Results of the Mission PDR analysis show that the purity with the current spectra recovery algorithms is limited strongly by the background, in particular by the out-of-field straylight and the Zodiacal light levels. The estimated performance at this point in time is shown in Table 10. It is expected that improvements on the ground segment processing, both in terms of straylight correction and spectra extraction will ultimately bring the requirement close to compliance. The choice of each pointing area in the Euclid survey design will be optimized also to limit the effect of out of field straylight. However it highlights the importance to maintain the focus on the straylight control both in the design and in the cleanliness and contamination which are a significant driver.

**Table 10 Spectroscopic performance**

| NISP-S Performance | | |
|---|---|---|
| Purity | 80% | 72% |
| Completeness | 45% | 52% |

In summary, the assessment of mission performance based on the as-designed system at Mission PDR shows robust margins on both image quality and radiometric levels providing a good basis to proceed into the implementation in Phase C/D. The marginal performance on purity is expected to be recovered once the maturity of ground processing algorithms progresses, however it reinforces the focus on straylight and cleanliness control already in place for the mission.

## 6. PROGRAMMATIC STATUS

Following mission approval in 2012, Euclid spacecraft has started the implementation phase at the end of 2012 with the PLM development with Airbus Defence and Space of Toulouse (F). Six months later, in July 2013, the industrial Prime contractor, Thales Alenia Space Italia of Torino (I), was selected. At the time of writing all the SVM and PLM subcontractors are actively progressing in their development. The industrial consortium successfully passed the preliminary design review in July 2015. The instruments teams, VIS led by UCL-Mullard Space Science Laboratory of Holmbury St. Mary (UK) and NISP led by CNES and the Laboratoire d'Astrophysique de Marseille (F) have completed the instruments preliminary design reviews in the first half of 2014 and are now well advanced in Phase C/D. The ground segment definition has also advanced consistently with the programme needs: the more classical OGS is currently undergoing to a Requirements Review, while the SGS, requiring more development started earlier and successfully passed its System Requirements Review in April 2015. A mission level Preliminary Design Review was successfully held in October 2015. All the mission elements advancement and the mission performance were assessed and considered fulfilling the scientific objectives within the programmatic constraints. The mission phase C/D is now in progress. The instruments Critical Design Reviews (CDR) are both expected to be concluded before the end of 2016. The spacecraft CDR is now planned for the first half of 2018 and Ground Segment Design review will be held in early 2018. The mission level CDR will be held in mid-2018. The schedule foresees the delivery of both instruments for integration into the PLM (optical part and cold electronic) and into the SVM (warm electronics) in the first half of 2018. At this point the flight system integration and test will start and will be carried out through the 2019 and 2020. The flight acceptance review is planned for the mid of 2020. The launch is currently planned for the end of 2020.